\documentclass[aps,prc,twocolumn,superscriptaddress,showpacs]{revtex4}
\usepackage{graphicx}
\usepackage{latexsym}
\usepackage{bm}
\usepackage{amsmath}

\usepackage{color}

\begin{document}

\title{\bf The Cronin Effect, Quantum Evolution and the 
Color Glass Condensate}

\author{Jamal Jalilian-Marian}
\affiliation{Physics Department, Brookhaven National Laboratory, Upton,
NY 11973, USA}

\author{Yasushi Nara}
\affiliation{%
Department of Physics, University of Arizona, Tucson, AZ 85721, USA
}

\author{Raju Venugopalan}
\affiliation{Physics Department \& RIKEN-BNL Research Center, Brookhaven 
National Laboratory, Upton, NY 11973, USA}

\date{\today}

\begin{abstract}
We show that the numerical solution of the classical $SU(3)$
Yang-Mills equations of motion in the McLerran-Venugopalan model for
gluon production in central heavy ion collisions leads to a suppression
at low $p_t$ and an enhancement at the intermediate $p_t$ region as
compared to peripheral heavy ion and pp collisions at the same
energy. Our results are compared to previous, Color Glass Condensate
inspired calculations of gluon production in heavy ion collisions.  We
revisit the predictions of the Color Glass Condensate model for $pA$
($dA$) collisions in Leading Order and show that quantum evolution--in
particular the phenomenon of geometric scaling and change of anomalous
dimensions--preserves the Cronin enhancement of $pA$ cross section
(when normalized to the leading twist term)
 in the Leading Order approximation even though the $p_t$ spectrum can
change. We comment on the case when gluon radiation is included.
\end{abstract}

\pacs{xxx}

\maketitle

\section{Introduction}

The Relativistic Heavy Ion Collider (RHIC) has opened a new frontier in
high energy heavy ion collisions. The first data from RHIC which showed a 
large suppression in the ratio of produced hadrons in $AA$ and $pp$ 
collisions \cite{rhic} has created much excitement in the heavy ion community.
It has led to intense theoretical and experimental work in order to 
understand and characterize the outcome. More recently, hadron spectra 
in $dA$ collisions at RHIC were measured \cite{rhic:dA}
in order to clarify the role of and distinguish
between initial state and final state (plasma) effects.
Whether the observed suppression of hadronic spectra~\cite{rhic:raa}
 and disappearance of back to back jets~\cite{star:btob}
in central Au+Au collisions,
 as well the large elliptic flow~\cite{rhic:v2}, in heavy ion collisions at 
RHIC is due to the Quark Gluon Plasma is still in need of further 
experimental investigation. Nevertheless, the recent results from the 
$dA$ collisions at RHIC and the apparent lack of strong initial state 
effects in mid--rapidity and at high $p_t$ appear to necessitate the presence 
of final state interactions in the partonic matter created in 
heavy ion collisions at RHIC~\cite{eloss}.

Even though the Cronin effect \cite{cronin} (the observation, at fixed 
target experiments, that the ratio of $pA$ and $pp$ cross sections,
scaled by the number of collisions, is above unity at some 
intermediate $p_t$ while below one at low $p_t$) is likely small in high
energy heavy ion collisions as compared to parton energy loss effects, 
it is one of the two main nuclear effects, along with shadowing, expected 
in high energy $pA$ ($dA$) collisions. Since $pA$ collisions were
proposed in order to clarify the role of initial and final state (medium) 
effects in high energy heavy ion collisions, it is extremely important
to have a firm understanding of the physics of shadowing and the Cronin effect
in order to have a precise understanding of the role of parton 
energy loss effects in heavy ion collisions.

The Cronin effect in high energy $pA$ and $AA$ 
collisions has been the subject of renewed theoretical 
interest recently \cite{adjjm,fgjjm,pa}.
In this brief letter, we show that the numerical solution of the 
classical Yang-Mills equations of motion \cite{KNV1,KNV2,lappi} in 
the McLerran-Venugopalan model \cite{mv} does indeed include the Cronin 
effect. We point out the differences between this numerical approach and 
other saturation
inspired models \cite{klm,lt} which led to the absence of the Cronin effect 
in these models. We re-visit some of our earlier results for $pA$ 
collisions and show that quantum evolution and the so called geometric 
scaling phenomenon \cite{IIM,gs} (plus change of anomalous dimension) 
preserves the Cronin enhancement even though its magnitude and 
peak may change. 

\section{The Cronin effect in $AA$ Collisions}

In the McLerran-Venugopalan model, the classical Yang--Mills equations
of motion are solved in the presence of external sources of color charge. 
These color charges can be thought of as the high $x$ quarks and gluons 
in the wavefunction of a nucleus and are Gaussian distributed with a 
characteristic scale $\Lambda_s$. (In practice, $\Lambda_s\approx Q_s$, 
the saturation scale.) In principle, $\Lambda_s$ can be determined from 
nuclear gluon distributions. It is an external parameter in the calculations 
described here. Physical quantities are computed by averaging over the 
Gaussian distributed color charges. 
The details of these computations for the real time gluodynamics 
of nuclear collisions can be found in Refs.~\cite{KNV1,KNV2,lappi}.

Briefly, the numerical lattice calculations in Ref.~\cite{KNV2}
impose color neutrality condition at the nucleon level
and realistic nuclear density profiles.
In our computations, we first solve 
the classical Yang-Mills equations on the lattice
for the two nuclei  before the collision.
In the radiation gauge ($x^+ A^- + x^- A^+ = 0$),
the initial conditions of gauge fields $A^{\mu}$
for nucleus-nucleus collisions at $\tau=0$
can be obtained by matching the solutions in the space-like and 
time-like regions. Requiring 
that the gauge fields be regular at $\tau=0$,
$D_{\mu i} F^{\mu i} = 0 $ and $D_{\mu +} F^{\mu +} = J^+$
for $x^-,x^+ \to 0$ gives the boundary conditions at $\tau=0$:
\begin{eqnarray}
 A^i(0,x_T) &=& A^i_1(0,x_T) + A^i_2(0,x_T),\quad (i=x,y), \nonumber\\
 A^{\pm}(0,x_T) &=& \pm x^{\pm}{i\over 2}
            [A^i_1(0,x_T),A^i_2(0,x_T)]\, .\nonumber
\end{eqnarray}
Here, $A^i_{1,2}$ are the pure gauge fields for two incoming nuclei.
Using these initial conditions, 
classical Yang-Mills equations are solved, assuming boost invariance,
 on a 2-dimensional
lattice.
The definition of the number
distribution in the non-perturbative region is discussed in
detail in \cite{KNV1}. In the transverse Coulomb gauge 
$\nabla_{\perp}\cdot A=0$, it is  
\begin{equation}
N(k) =
 \sqrt{\langle
|\phi(k)^2|\rangle\langle|\pi(k)|^2\rangle},
\nonumber
\label{eq:coulomb} 
\end{equation} 
where $\phi(k)$ and $\pi(k)$ correspond to the potential and kinetic terms
in the Hamiltonian respectively.

\begin{figure}[htp]
\includegraphics[width=3.3in]{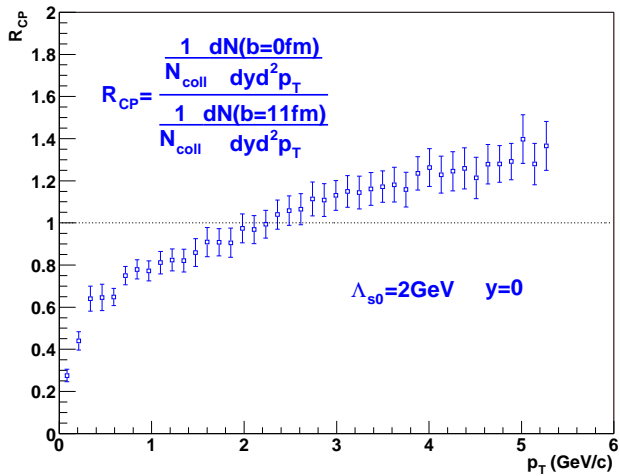}
\caption{$R_{CP}$ from the McLerran Venugopalan model for an SU(3) gauge 
theory. Here  $\Lambda_{s0}=2$ GeV, where
$\Lambda^2_{s0}$ is the color charge squared per unit transverse
area in the center of each nucleus. (The value of $\Lambda_s$ averaged 
over the entire nucleus is smaller $\sim 1.4$ GeV.)
This result is obtained for a $256\times256$ lattice.
}
\label{fig:R_cp}
\end{figure}

In Fig.~\ref{fig:R_cp}, we plot $R_{CP}$, the ratio of produced gluons in 
head-on ($b=0$ fm) and in peripheral ($b=11$ fm) $Au$-$Au$ collisions 
normalized by the number of collisions $N_{\mathrm{coll}}$, for an SU(3) 
gauge theory. $N_{\mathrm{coll}}$ is computed self-consistently in our 
framework and agrees, for instance, with Ref.~\cite{kln}.  
The ratio $R_{CP}$ is below unity at low $p_t$ and above unity
at intermediate $p_t$. This shows that the original McLerran-Venugopalan 
model indeed has the Cronin enhancement. One can in principle show that 
this ratio goes to unity at high $p_t$, but this is numerically intensive and 
not shown here. 

\begin{figure}[htp]
\includegraphics[width=3.3in]{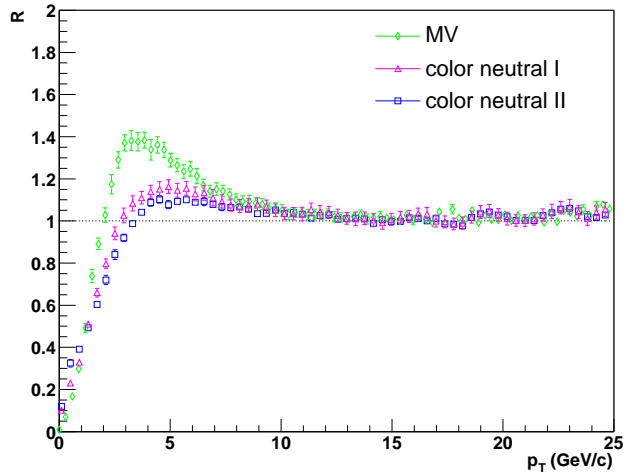}
\caption{$R_{AA}$ for an SU(2) 
gauge theory. $R_{AA}$ is the ratio of the $p_t$ distribution 
of gluons for $Au$-$Au$ collisions ($\Lambda_{s0}=2$ GeV) divided by 
$p$-$p$ collisions ($\Lambda_{s0}=0.2$ GeV) and 
normalized by the ratio of their asymptotic 
values at large $p_t$. Here MV denotes the McLerran-Venugopalan model with 
color neutrality imposed globally; Color Neutral I \& II impose 
color neutrality on each configuration at the nucleon level-see text for 
discussion.  
}
\label{fig:R_su2cp}
\end{figure}

\begin{figure}[htp]
\includegraphics[width=2.5in]{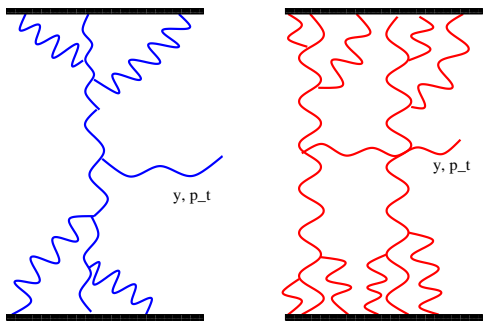}
\caption{Schematic diagram of mono-jet gluon production amplitude in 
$k_t$ factorized form 
(left figure) vs the non-factorized contributions arising from 
solutions of classical Yang-Mills equations (right figure).}
\label{fig:monovsclas}
\end{figure}

Instead, for simplicity, we show in Fig.~\ref{fig:R_su2cp}, results from a 
computation of the real time evolution of an SU(2) gauge theory. There is 
no qualitative difference between the SU(2) case and the SU(3) case. Our 
results for the SU(2) gauge theory are obtained for 
larger (512$\times$512) lattices relative to the (256$\times$256) lattices 
in the SU(3) case. Plotted in 
Fig.~\ref{fig:R_su2cp}, 
is the ratio of the $p_t$ distribution of gluons from $Au$-$Au$
collisions (with $\Lambda_{s0}=2$ GeV) to that in $p$-$p$ collisions,
normalized to the ratio of their asymptotic values ($N_{\rm coll}$) at
high $p_t$. The $p$-$p$ results here are taken to be the lattice
results for the very small value of $\Lambda_{s0}=0.2$ GeV. The
lattice result in the latter case is equivalent to the perturbative
tree level result up to very small $p_t$'s.  The reason $R_{AA}$
deviates from unity is due to the multiple scattering (``Cronin'')
effect illustrated in Fig.~\ref{fig:monovsclas}-this point will be
discussed further shortly.  The three curves in Fig.~\ref{fig:R_su2cp}
correspond to the following: MV denotes the McLerran-Venugopalan model
with color neutrality imposed only globally over the entire nucleus in
each configuration.
Color Neutral I \& II correspond to the more stringent condition where 
color neutrality is imposed on the scale of a nucleon. The former corresponds 
to the case where monopole component of the nucleon color charge is subtracted 
while in the latter case, both monopole and dipole components are subtracted. 
As has been noted previously in Ref.~\cite{KNV2}, the effect of these more 
stringent color neutrality conditions is to induce a power law dependence 
$p_t^n$ at low $p_t$ for the correlator of color charges in momentum space. 
This dependence, albeit non-perturbative in origin, is similar to that 
induced by perturbative color neutrality 
which arises from the color screening of saturated gluons generated in the 
quantum 
evolution~\cite{IIM}. The softening of the Cronin enhancement can therefore 
be understood as resulting from color screening or ``shadowing'' of the 
initial gluon distributions. 

In a nice paper, Kovchegov and Mueller have shown that the choice of 
gauge can determine whether interactions are ``initial state'' or 
``final state'' effects~\cite{kovmu}. In our numerical computations 
(carried out in $A^\tau=0$ gauge)
initial state as well as some final state 
interactions are included. We qualify the latter because  
due to the strong expansion of 
the system with time, gluon field strengths decrease very quickly. 
High  $p_t$ modes ($p_t > \Lambda_{s0}$) in particular are linearized at very 
early times and therefore, suffer no further final state 
interactions. On the other hand, soft modes still interact with each 
other for a long time as shown in Ref.~\cite{KNV3}.
Therefore, the lattice result of $R_{CP}>1$ in Fig.~\ref{fig:R_cp} 
can be regarded as the Cronin enhancement resulting from these re-scatterings.
The interaction between hard and soft modes for occupation numbers $f\leq 1$ 
remains an outstanding problem. Naively, these interactions are of higher 
order in the classical approach but at late times, with the classical 
approach breaking down, they 
become competitive. Despite recent progress~\cite{BMSS}, a full treatment 
of this problem is lacking.

\begin{figure}[htp]
\includegraphics[width=2.5in]{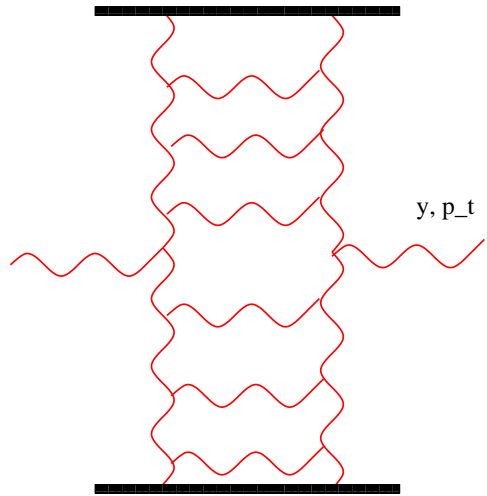}
\caption{Mono-jet production cross section in $k_t$ factorized form.}
\label{fig:monojet}
\end{figure}

We now compare our treatment to other approaches in saturation based models. 
Since there is no known complete 
{\it analytical} solution for the collision
of two nuclei \cite{kmw,yk}, a {\it simplified} model was used in \cite{kln}. In this ``$k_t$ factorized'' approach, the 
cross section for gluon productions is given by
\begin{eqnarray}
E{d\sigma \over d^3p} = {4\pi N_c \over N_c^2 -1} {1 \over p_t^2}
\int^{p_t^2} dk_t^2 \alpha_s\, \phi_A(x_1,k_t^2) \phi_A(x_2,(p_t-k_t)^2)
\nonumber
\end{eqnarray}
where $\phi_A(x,p_t)$ is the unintegrated gluon distribution function of
a nucleus
\begin{eqnarray}
G_A(x,Q^2)\equiv \int^{Q^2} dk_t^2 \, \phi_A(x,k_t).
\nonumber
\end{eqnarray}
This equation describes mono-jet production in the $k_t$ factorized form
and can be visualized as in Fig. \ref{fig:monojet}. Radiation of additional
jets are down by powers of $\alpha_s$. 
The connection to gluon saturation and the Color Glass Condensate 
is phenomenological and comes from the assumed form of the unintegrated 
gluon distribution $\phi_A(x,k_t)$ in the transverse momentum region 
$k_t^2 \ll Q_s^2$. In the perturbative region where $ k_t^2 \gg Q_s^2$ 
\begin{equation}
\phi_A(x_1,k_t^2) \sim 1/k_t^2 
\end{equation}
while in the saturation region where $ k_t^2 \ll Q_s^2 $
\begin{equation}
\phi_A(x_1,k_t^2) \sim 1/\alpha_s.
\end{equation}
In \cite{kln} these two asymptotic forms of the unintegrated gluon
distribution are then matched at $Q_s$. In this approach, there is only 
a single scattering in the $k_t > Q_s$ region and multiple 
scatterings at scales just above $Q_s$ are not included. Since 
the non-perturbative input in the form of the ``gluon liberation factor''
is included from the numerical lattice simulations~\cite{KNV1}, this 
approach likely includes phenomenologically 
the physics of the CGC for global observables 
such as the centrality and rapidity dependence of observables (further 
discussed in the next paragraph). 

The same $k_t$-factorized formalism is considered in Ref.~\cite{klm}. The authors 
however additionally consider the effects, at moderate $p_t$'s, due to the change in the 
unintegrated gluon distributions arising from quantum evolution. In particular, they take into 
account the change in the distributions due to the modification of the anomalous dimensions 
in the evolution. This change leads to a scaling with $N_{\rm part}$ as opposed to $N_{\rm coll}$ scaling. 
(Note that these modifications do not affect the prior results of Ref.~\cite{kln} since the multiplicities, 
at the $p_t$'s at which they occur, are rather small.) However, 
the Cronin enhancement for $p_t$ distributions is missed in \cite{kln,klm}. 
On the other hand, in our numerical computations, all 
re-scatterings are included at
the classical level leading to the Cronin enhancement. Note though that 
quantum evolution is absent in our lattice calculations-its effects 
are only  implicitly included through the magnitude of the saturation scale. 

In addition to their importance for the Cronin effect, the 
non-factorized contributions illustrated in Fig.~\ref{fig:monovsclas} 
likely play a significant role in determining the total gluon multiplicity 
as well. This can be understood by comparing the 
numerical results for the gluon liberation coefficient $f_N$~\cite{KNV1,lappi} 
which give $f_N=0.3-0.5$, in contrast to analytical results for this 
coefficient that include only the factorized contributions and is roughly 
 3 to 4 times larger~\cite{yk}.

A possible way of including this enhancement in the $k_t$ factorized form 
through quantum evolution is discussed in \cite{bkw}. A similar 
approximation of $\phi_A (x,k_t)$ in the context
of $pA$ collisions \cite{lt} also leads to a lack of Cronin enhancement 
in the cross section. It would be interesting to include quantum evolution 
effects in addition to the Cronin effect in our numerical work to quantify 
the importance of the energy loss contribution. This work is in progress 
and will be reported shortly~\cite{hirano}.
Below, we consider $pA$ scattering and explicitly
show that inclusion of higher order scatterings leads to the Cronin effect.

\section{The Cronin effect in $pA$ Collisions} 

Even though the Cronin effect in a high energy heavy ion collision is
quite small at high $p_t$ as compared to the parton energy loss 
effects, it is quite important in a proton (deuteron) nucleus collision
since energy loss effects in the hot medium  produced in a heavy ion collision,
are absent in a $pA$ collision. In this section, we re-visit the 
Color Glass Condensate results for a quark or gluon scattering on a 
nucleus \cite{adjjm} and show that the recent phenomenon of geometric 
scaling and changing of anomalous dimension
does not affect our conclusions as long as we divide our $pA$ cross section
by its leading twist limit (rather than $pp$).

Scattering of a quark or gluon on a nucleus using the Color Glass
Condensate formalism and its relation to DIS was discussed in 
\cite{adjjm,fgjjm}. It was shown that the scattering cross section 
is given by
\begin{eqnarray}
{d\sigma^{qA\to qX} \over d^2 p_t\,dp^-\,d^2b_t}& & = {1 \over (2\pi)^2}
\,\delta(q^--p^-) \nonumber\\
&\times&
  \int d^2 r_t \, e^{i p_t \cdot r_t} \, [2 - \sigma_{\rm {dipole}}(x,r_t,b_t)] 
\label{eq:qAcs}
\end{eqnarray}
with 
\begin{eqnarray}
\sigma_{\rm {dipole}} (x,r_t,b_t) \equiv {1\over N_c} Tr\,
\left<
1 - V(b_t +\frac{r_t}{2}) V^\dagger(b_t -\frac{r_t}{2})
\right>_\rho
\label{eq:sigdipole}
\end{eqnarray}
where $q^- (p^-)$ is the longitudinal momentum of the incoming (outgoing)
quark, with a similar equation for gluon scattering. This is the multiple
scattering generalization of quark gluon scattering in pQCD and
unlike the leading twist (single scattering) result, is finite as
$p_t \rightarrow 0$ due to higher twist effects. In addition, defining
\begin{eqnarray}
\gamma(x,p_t,b_t)\equiv \int d^2 r_t \, e^{i p_t \cdot r_t} \, 
[2 - \sigma_{\rm{dipole}}(x,r_t,b_t)] 
\label{eq:gammadef}
\end{eqnarray}
and using the fact that $\sigma_{\rm{dipole}}(x,r_t=0,b_t)=0$, we see that 
$\gamma$ satisfies the following sum rule
\begin{eqnarray}
\int d^2p_t \gamma(x,p_t,b_t) = 2\,(2\pi)^2 \,,
\label{eq:sumrule}
\end{eqnarray}
for fixed $b_t$. It is clear from this sum rule that if the cross section in (\ref{eq:qAcs})
is suppressed at low $p_t$, it must be enhanced at high $p_t$ in order
to compensate for the low $p_t$ suppression so that the sum rule holds 
\footnote{In the first paper of \cite{adjjm}, a careless
wording of the paragraph after eq. $24$ gives the impression that
there is no Cronin effect. This is not the case and the $qA$ scattering 
cross section, first derived there, does have the Cronin enhancement.}. 
The Cronin peak moves to higher $p_t$ as one goes to higher energy 
(or more forward rapidities as well as more central collisions).

The effects of quantum evolution on our classical results can be
investigated using the Renormalization Group equations \cite{rg}, and
in particular the Balitsky-Kovchegov (BK) \cite{bk} equation for the 
dipole cross section in the large $N_c$ limit. It is straightforward 
to apply the BK equation to the $qA$ (or $gA$) scattering cross section
in order to prove that quantum evolution preserves the sum rule in
(\ref{eq:sumrule}). For our purposes,
it suffices to notice that $\sigma_{\rm{dipole}}(x,r_t=0,b_t) = 0$ at
any $x$ (energy) so that the sum rule is preserved by quantum evolution
and therefore, the Cronin enhancement is still present. Nevertheless,
the $p_t$ spectra will look different at different energies since the
location of the Cronin peak as well as its magnitude changes as one 
goes to higher energies (see for example, Figure $4$ in the last paper 
of \cite{fgjjm}.). In addition, the change of the anomalous dimension 
$\gamma$ could modify the magnitude of the Cronin effect. Nevertheless, 
since the BK equation preserves the sum rule in (\ref{eq:sumrule}) at
the partonic level, this modification in the high $p_t$ region will 
be compensated by an analogous modification in the low $p_t$ region 
so that the sum rule is not violated.

One should keep in mind that at RHIC and for mid rapidity and high 
$p_t$ processes (such as the suppression of hadron yields) the effective 
$x$ of the partons is quite large. For example, for $p_t \sim 5-10$ GeV 
in mid rapidity, the $x$ range is $\sim 0.05 - 0.1$. The only 
potential evidence for gluon saturation and geometrical scaling in hadrons is 
from HERA where saturation models work very well only for $x < 0.01$ and 
fail at higher $x$. In the case of nuclei, this effective $x$ may be 
slightly higher but can not be too much higher since otherwise, strong gluon 
shadowing would manifest itself in the $F_2$ structure function which
shows no strong shadowing effects at the $x$ range $0.05 - 0.1$ and as a 
matter of fact, shows anti shadowing! The Color Glass Condensate model
is an effective theory of QCD at small $x$ only and likely breaks down 
at the high $x$'s relevant for high $p_t$ processes in mid rapidity
at RHIC. (This situation will improve as one goes to higher collision 
energies or stays in the low $p_t$ region.) 

Moving to more forward rapidities (in the projectile fragmentation
region) will make applications of the Color Glass Condensate model 
more reliable since smaller values of $x$ in the target are probed.  
The  saturation scale is larger, rendering weak coupling
methods more reliable. In addition, the contribution of high $x$ region to the
cross section will be less important than that in the mid rapidity region.
Whether the most forward rapidities accessible at RHIC are large enough
for the physics of gluon saturation to be the dominant physics remains 
to be seen.

\section{Summary}

We considered in this note 
gluon production in heavy ion collisions using the
numerical simulations of the Color Glass Condensate. We showed that 
the ratio of central to peripheral cross sections (or equivalently, 
the ratio of central to pp cross sections), 
normalized by the number of collisions, shows the 
Cronin enhancement at high $p_t$ as well the suppression at low $p_t$.
We discussed other gluon saturation and 
Color Glass Condensate inspired models and commented
on the absence of the Cronin effect in these models.
We also considered $pA$ collisions and showed that, at the partonic level 
and in leading 
Order in $\alpha_s$, quantum evolution with energy and the change of 
anomalous dimension preserves the Cronin enhancement
(when normalized to its leading twist term rather than $pp$)
 due to a sum rule 
satisfied by the dipole cross section even though the magnitude and 
location of Cronin peak is energy dependent.
When divided by the $pp$ cross section as done experimentally,
our ratio will have suppression at high $p_t$ in agreement
with the results of Ref.~\cite{KKT}.

After this work was completed, we were made aware of similar work by 
Kharzeev, Kovchegov and Tuchin~\cite{KKT}. Although there is some overlap 
in our discussion of the pA case (their focus here being on 
gluon production), they do not explicitly 
consider the Cronin effect for AA collisions as we have.

\begin{acknowledgments}
J.J-M. and Y.N. would like to thank the Institute for Nuclear Theory
at the University of Washington for hospitality and the Department 
of Energy for partial support while this work was being completed. 
J. J-M. \& R.V. are supported by the U.S.\ Department of Energy under 
Contract No.\ DE-AC02-98CH10886 and J.J-M. is supported 
in part by a Program Development Fund Grant from Brookhaven Science Associates.
R.V. thanks the RIKEN-BNL Research Center at BNL for continued support. 
R.V. and Y.N. thank Alex Krasnitz for discussions.
Y.N.'s research is supported by the DOE under Contract No.~DE-FG03-93ER40792.
\end{acknowledgments}

\end{document}